\documentclass{mn2e}
\usepackage{graphicx}
\usepackage{amssymb}
\usepackage{float}

\def\ltsima{$\; \buildrel < \over \sim \;$}
\def\simlt{\lower.5ex\hbox{\ltsima}}
\def\gtsima{$\; \buildrel > \over \sim \;$}
\def\simgt{\lower.5ex\hbox{\gtsima}}
\def\kpc{{\rm\,kpc}}

\def\msun{{\rm\,M_\odot}}

\newcommand{\fmmm}[1]{\mbox{$#1$}}
\newcommand{\scnd}{\mbox{\fmmm{''}\hskip-0.3em .}}

\def\s{\ifmmode \widetilde \else \~\fi}
\def\={\overline}

\def\spose#1{\hbox to 0pt{#1\hss}}

\def\lta{\mathrel{\spose{\lower 3pt\hbox{$\mathchar"218$}}
     \raise 2.0pt\hbox{$\mathchar"13C$}}}
\def\gta{\mathrel{\spose{\lower 3pt\hbox{$\mathchar"218$}}
     \raise 2.0pt\hbox{$\mathchar"13E$}}}
\def\Dt{\spose{\raise 1.5ex\hbox{\hskip3pt$\mathchar"201$}}}    
\def\dt{\spose{\raise 1.0ex\hbox{\hskip2pt$\mathchar"201$}}}    

\def\dotsfill{\leaders\hbox to 1em{\hss.\hss}\hfill}

\title[The Andromeda Stream]{The three dimensional structure of the giant stellar stream in Andromeda}

\author[McConnachie et al.]
     {A. W. McConnachie${^1}$, M. J. Irwin${^1}$, R. A. Ibata${^2}$, A. M. N. Ferguson${^3}$,\newauthor
G. F. Lewis ${^4}$, N. Tanvir${^5}$\\  
${^1}$ Institute of Astronomy, Madingley Road, Cambridge, CB3 0HA, U.K.\\
${^2}$ Observatoire de Strasbourg, 11, rue de l'Universite, F-67000, Strasbourg, France\\
${^3}$ Max-Planck-Institut f\"{u}r Astrophysik, Karl-Schwarzschild-Str. 1, Postfach 1317, D-85741 Garching, Germany\\
${^4}$ Institute of Astronomy, School of Physics, A29, University of Sydney, NSW 2006, Australia\\
${^5}$ Physical Sciences, Univ. of Hertfordshire,Hatfield, AL10 9AB, U.K.\\}

\begin{document}

\maketitle

\begin{abstract}
The wide-field CCD camera at the  CFH telescope was used to survey the
giant  stellar stream in  the Andromeda  galaxy, resolving  stars down
the red giant  branch in M31 to $I \simeq 25$, a magnitude deeper than our previous INT survey of this galaxy and extending $1^{\circ}$ further out. The stream is seen
to extend out to the south-east of  M31 as far as we have surveyed
(some  $4.5^{\circ}$,  corresponding  to  a projected  distance  $\sim60
\kpc$). It is a linear structure in projection, and the eastern edge
of the stream presents a sharp boundary in star counts suggesting that
it remains a coherent structure.  By analysing the luminosity function
of the metal  rich component of the stream we find that, at the  furthest extent of our
survey, the stream  is $100\kpc$ further away along  the line of sight
than M31. It can then be traced to a point on the north-western side of the galaxy where  it is some 30  kpc in front  of M31, at which  point the
stream turns away from our survey area.

\end{abstract}


\section{Introduction}

Stellar  streams  are the  remnants  of  structures  that have  become
perturbed or  disrupted due to  the action of gravitational  tides. In
the  hierarchical formation  model  of galaxy  haloes, smaller  systems
merge first and  form the larger galaxies we see  today (White \& Rees
1978).  As  such, these  structures  are  expected  to be  common  and
examination of  their properties gives valuable  information about the
galactic formation  process. In the outer regions  of haloes, dynamical
times are  sufficently long to  leave many structures  detectable over
much of  the age of  the universe (eg. Johnston, Hernquist \& Bolte 1996) and  so these streams can  trace the
left-overs of  even ancient accretions. Indeed,  under some favourable
conditions the galaxy potential  can leave massive streams essentially
intact for all cosmic time.

Within our Local  Group there is a lot  of observational evidence that
galaxy  mergers  are  an   on-going  process.  The  discovery  of  the
Sagitarius dwarf galaxy by Ibata, Gilmore \& Irwin (1994) revealed that our own Galaxy is
undergoing strong interactions with its companions. Further observations detected a tidal stream associated with this galaxy's disruption (eg. Ibata et al. 2001a). Indeed, several of
our Galaxy's globular clusters have been found to have originally been
associated with  this dwarf galaxy  (Ibata, Gilmore \& Irwin 1994; Bellazzini, Ferraro  \& Ibata
2003). One  of the most favoured explanations  for the formation
of  the thick  disk  is that  it is  due  to a  violent merging  event
(eg. Quinn, Hernquist \& Fullager 1993, Schwarzkopf \& Dettmar 2000, Feltzing et al. 2003). Disentangling
the merger history of galaxies is thus of prime importance and streams
are ideal tracers of this process.

The discovery  of a giant stellar  stream in the outer  regions of our
neighbouring   giant  galaxy   M31   was  reported   in  two   earlier
contributions in this series (Ibata et al. 2001b, Ferguson et al. 2002).
It is an apparently vast structure  and our initial INT WFC survey did
not go out far enough to find  its true extent.  Here we report on our
findings from  a deeper  survey using the  CFH12K camera that probes further out and to fainter magnitudes than we had previously surveyed.

\section{Observations}

During the  nights of August  23, September 13-14 and  September 17-18
2001,  fourteen  fields were  oberved  with  the 12k$\times$8k  CFH12K
camera, at the prime focus of the Canada-France-Hawaii telescope.  The
observations  were kindly  taken by  CFH Telescope  staff,  in service
mode.   Conditions  were generally photometric,  with  good  seeing of  $\approx
0\scnd8$.  Three sets of Mould V-band exposures and three Mould I-band exposures,
each of  545s duration,  were secured per field. A negligible colour correction is involved to convert these filters to the standard Johnson-Kron-Cousins system \footnote{http://www.cfht.hawaii.edu}. The  data were  preprocessed by
CFH  staff with  the  CFH  pipeline software,  to  correct for  bias
offset, flat-fielding  and fringing.  We then applied  the same object
detection,   classification,   photometry   and   catalog   generation
algorithms  as we  successfully used  before on  our  INT-based survey
(Irwin \& Lewis 2001).

\begin{figure}
\includegraphics[angle=270,width=8cm]{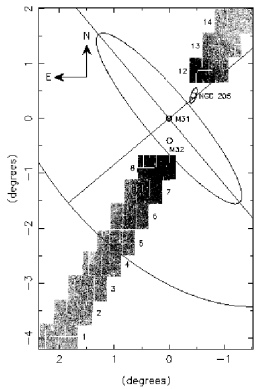}
\caption{The location of the  CFH12K fields used in this paper. The $2^{\circ}$ radius ellipse  marks the outer boundary of the
optical disk  of M31, while  the curve is  a segment of  a $4^{\circ}$
radius  ellipse of  semi-major axis  $\simeq 55  \kpc$  and flattening
$0.6$ (this corresponds to the  original limit of our INT survey). The
CFH field number  is indicated either to the right or  to the left of
each field. Fields 9-11 are heavily crowded and will be the subject of
a subsequent contribution.}
\label{map} 
\end{figure}

The fields were  selected to lie along the  previously detected region
of the  Andromeda stream, and we  extrapolated along this  stream in a
straight line  $\sim 3$ degrees further  out to the  south-east and to
the north-west.  The arrangement  of the survey regions presented here
is  displayed  in  Figure  \ref{map}, and the RA and Dec of fields 1--8 and 12--14 are listed in Table 1.  The  photometry  of  the  three
heavily-crowded fields  9--11, which lie  close to the center  of M31,
will be presented elsewhere. Field 14 is used as a reference field, as no stream component was found there.

\section{Stellar populations of the stream}

\begin{figure*}
\includegraphics[angle=270,width=10cm]{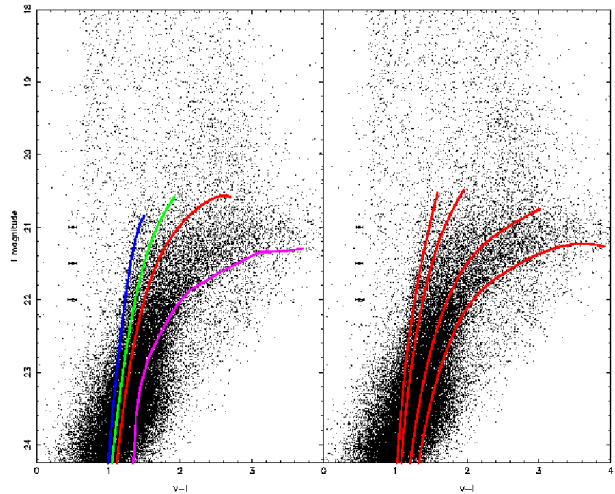}
\caption{Colour-magnitude diagrams  of field 6. Error bars show typical uncertainties in the photometry at $I = 21.0, 21.5$ and $22.0$. On the panel on the left we  have overlayed 4 well studied globular clusters sequences of  different metallicities. From  left to right,  these are
NGC6397 ($[Fe/H] = -1.9$), NGC1851 ($[Fe/H] = -1.3$), 47Tuc ($[Fe/H] =
-0.7$) and NGC6553 ($[Fe/H] = -0.2$).  In the panel on the right are 4
VandenBerg tracks for a $0.8 \msun$ star with $[\alpha/Fe] = 0.3$. The
metallicities  that  these correspond  to,  from  left  to right,  are
$[Fe/H]  = -1.8$,  $[Fe/H]  = -1.3$,  $[Fe/H]  = -0.7$  and $[Fe/H]  =
-0.4$. For  all the  overlays, we have  assumed a distance  modulus of
24.47 and an average reddening of $E(B-V) = 0.07$ (following Schlegel et
al. 1998).  }
\label{cmds} 
\end{figure*}

A typical field from our survey (field 6) that illustrates the quality
of the CFHT  data is shown in Figure \ref{cmds}. This  field is one of
several that  overlaps the INT survey  region, although the CFHT data extends a full magnitude deeper than our previous survey. The overlap enables us  to make an
external comparison  of the photometric systems  and calibration.  The
internal  overlaps  between  (most)  CFH fields  and  their  external
overlap with some of the INT data allowed us to update the calibration
of   two  CFH   fields   (2,3)  that   were   taken  under   slightly
non-photometric  conditions. After  these adjustments  the photometric
calibration systematic  errors are at the level  of $\approx \pm$2$\%$
over  the entire  region surveyed.   Overlaid on  the colour-magnitude
diagram  are four  well-studied globular  cluster  sequences (NGC6397,
NGC1851,   47Tuc  and   NGC6553) from Da Costa \& Armandroff (1990) and Sagar et al. (1999) that span  a   range  of   different
metallicities  (left panel),  and also  four evolutionary  tracks (for
$\alpha$-enhanced  $0.8\msun$  stars) from  VandenBerg  et al.  (2000)
(right panel).   A wide  range of stellar  populations are  evident in
this field spanning the full range of metallicities represented by the
fiducial  sequences ($-0.2  \ge [Fe/H]  \ge -1.9$).   Inspection  of a
sequence  of  these diagrams  shows  evidence  for  a metal-poor  halo
population with a metallicity  spanning roughly $[Fe/H] \approx -2$ to
$[Fe/H]  \approx -0.7$  and, in  addition  to this,  a more  metal-rich
population  is evident  in  the  majority of  the  fields.  Using  the
fiducial sequences to define the metallicity distribution function we  can attribute  a mean  metallicity of
$[Fe/H] \approx -0.5$ to this component, with a dispersion of $\approx
-0.5$dex. This population is clearly seen to extend out to field
1  in  our  CFH  survey  data,  where it  still  remains  a  numerous
component:  around  1000 stream  stars  are  detected  in field  1  to
I$ = 23.5$.  This is  in contrast to the metal  poor halo component that
virtually   disappears   by   this  galactocentric   distance   ($\sim
4^{\circ}$).

The evolutionary tracks we have selected correspond to $[\alpha/Fe] = 0.3$ since CNO and other $\alpha$-process elements, which dominate the heavy element composition, are thought to be enhanced relative to solar for Galactic halo stars (eg. Wheeler, Sneden \& Truran 1989, Carretta, Gratton \& Sneden 2000). These tracks are seen to be  in good agreement
with the globular cluster sequences and our data. In particular, the I
magnitude of  the tip  of the  red giant branch  (TRGB) of  the tracks
appear to  match up well as  a function of metallicity.   We will make
use of  this property in  Section \ref{m31dist} in order  to determine
the relative distance of the  metal-rich population to the bulk of the
stellar population of M31.

\section{Three dimensional structure of the Andromeda Stream}

\subsection{Stellar profile across the stream}

\begin{figure}
\includegraphics[width=6.5cm,angle=270]{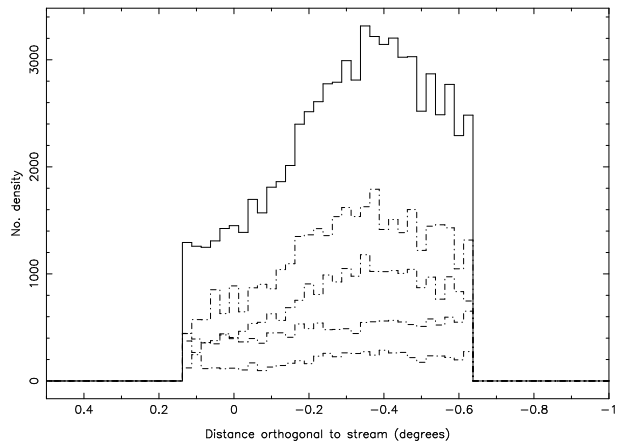}
\caption{ The distribution  of stars satisfying $20.5 <
{\rm I} < 22.5$ and ${\rm V-I} > 2$ orthogonal to the stream in fields
1-8. The shallowest histogram (dotted) is for fields 1--3, then for fields 4--5, 6--7, and finally 8. The solid line is the combined histogram for all these fields. The distance  displayed on  the abscissa  is  the perperdicular
distance from the line that is  parallel to our survey and that passes
through the centre  of M31. The counts increase  sharply from the left
hand (eastern) side, a feature  which is consistently seen  along the
stream.  }
\label{orth}
\end{figure}

In the  INT data, the stream  appeared as a  fairly straight structure
pointing away to the SE from  the center of the galaxy. With our extra
CFH data we  are now in a position to  better constrain the projected
profile of  the stream on the  sky.  A reference line  was chosen that
follows a path parallel to the direction of the CFH survey fields and
that  passes  through  the  centre  of M31.  The  density  of  sources
perpendicular  to this line  satisfying $20.5  < {\rm  I} <  22.5$ and
${\rm V-I} > 2$, for fields 1 to 8, is displayed as a solid line in Figure \ref{orth}. This figure demonstrates  that there is a rapid  rise in number counts
from the NE  side of the stream towards the SW  side, with the density
of   sources    increasing   by   a   factor   of    $\sim   2$   over
$0.5^\circ$. However, after the  peak, the distribution is almost flat
up to  the SW edge  of our survey  area.  This is consistent  with the
sharp edge to the stream seen in the INT survey, although at this time
we cannot  constrain  the behaviour  of  the SW  edge of  the
stream.   To investigate  the radial  dependency we  divided  the CFH
survey data into 4 bins (dotted lines). In Figure \ref{orth}, the shallowest histogram is for fields 1 -- 3, next is for fields 4--5, then 6--7 and finally 8.  The distribution of sources in each of these bins can be observed to have a similar shape and, in particular,
we find  that the position of  the peak of the  distribution varies by
less than  $0.2^\circ$ over  the $4^\circ$ line  between fields  1 and
8. However, the spatial profile of the stream beyond $d =  2.4^\circ$ is broader, showing that the stream of  width $\sim 0.5^\circ$
appears to fan out into a wider distribution in the outer regions.

Likewise, by separating the stars in  Figure \ref{orth} with $d < 0.9^\circ$, we find that   there   is   no    significant   difference in the shape of the histograms   with that for $d > 0.9^\circ$. We had expected that for $d < 0.9^\circ$ the contribution of the stream would be small compared to the halo, and so the histograms would be flatter in appearance. Instead, it would appear that the
stream contamination is still significant in the $d < 0.9^\circ$ region.

\begin{figure}
\includegraphics[width=6.5cm,angle=270]{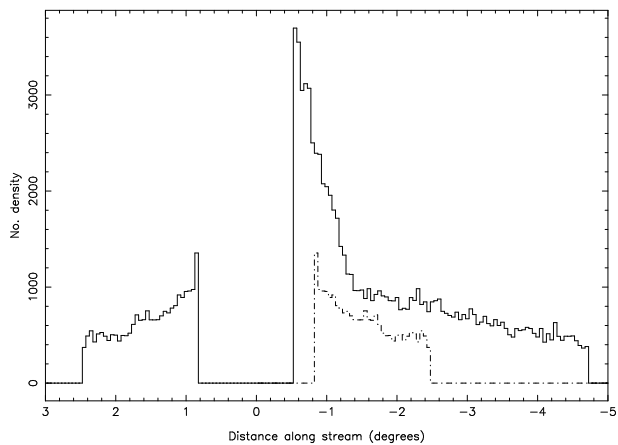}
\caption{The distribution  of stars along the path
of the stream, as defined in  the previous figure. The dotted line is a mirrored version of the data from fields 12 and 13, to illustrate the large excess of stars due to the stellar stream obvious in fields 1-8 for  a distance  $< -1.5^{\circ}$. The dip at $-2^{\circ}$ is due to the gap between fields 5 and
6.}. 
\label{parallel} 
\end{figure}

Figure \ref{parallel} shows how the number density of these stars varies along 
the path defined above (fields 9, 10 and 11 have not been used due to crowding 
problems). As expected, the star counts increase as we approach the centre of 
M31. However, there is an excess of stars clearly visible for $d < -1.5^\circ$, which is highlighted in Figure \ref{parallel} by comparison with the profile for $d > 0^\circ$ (its reflection about $ d = 0^\circ$ is shown as a dot-dashed line). The CFH fields were selected to highlight the stream which is clearly delineated in azimuth (Fergusson et al. 2002) thus we attribute this excess to the stream.  Although we believe stream to be present at $d > 0^\circ$ (Section 4.3), its stellar density by this point is much smaller compared to other M31 components.

\begin{figure*}
\includegraphics[angle=270,width=10cm]{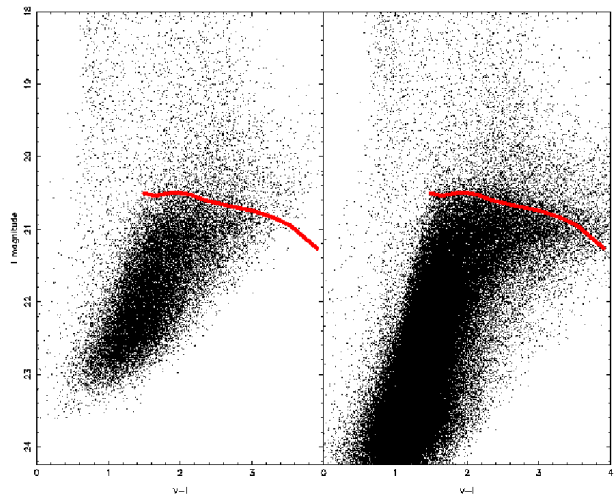}
\caption{On the left is  an inner field from our M31
INT  WFC  survey,  converted  to  V and  I  using  the  transformation
equations given in the text. The  location of  the tip  of the red  giant branch  (TRGB) is
readily seen.  Overlaid is the behaviour  of the tip as  a function of
metallicity  as given  by  the evolutionary  tracks  of Vandenberg  et
al.  (2000). Detailed  inspection shows  that the  tracks are  in good
agreement with  the data. On  the right is  field 8 from our  CFH12K
survey.  By  calibrating the  accuracy  of  the  Vandenberg tracks  in
predicting the magnitude of the TRGB  in our M31 field on the left, we
can  measure  the  distance  variation  between  the  stream-dominated
component ($V-I \gta 2.2$)  and the M31-dominated component ($V-I \lta
2.2$) in the  field on the right. This places  the stream component in
field 8 approximately at the same distance as the bulk of M31.}
\label{field76}
\end{figure*}

\subsection{Stream distance relative to M31}
\label{m31dist}

The I band magnitude of the tip  of the red giant branch (TRGB) is now
recognised as  a good standard  candle for metal  poor ($[Fe/H] < -0.7$)
old  ($> 2$   Gyrs)  stellar   populations  (eg.  Salaris   \&  Cassisi
1997). However, as  we go to redder, more  metal rich populations, the
luminosity decreases  due to  the effect of  increased opacity  in the
stellar atmosphere.  This is clearly observed in  our colour magnitude
diagrams. As such, in order to  gain a distance estimate of the stream
relative  to M31, we  need to  calibrate the  redder (${\rm  V-I} \gta
2.2$),  metal-rich stream  to the  bluer  (${\rm V-I}  \lta 2.2$)  M31
component and therefore  need to know how the  TRGB behaves a function
of colour.

Vandenberg et  al. (2000) present  evolutionary tracks for a  range of
stellar  masses   ($0.5  \le  \msun   \le  1.0$),  with  a   range  of
metallicities  ($-2.31 \le [Fe/H]  \le -0.30$).  A selection  of these
tracks are shown in Figure \ref{cmds}.  In the left  panel of Figure
\ref{field76}  we  show  the   predicted  location  of  the  TRGB  for
$0.8\msun$ stars with $[\alpha/Fe] = 0.3$ as a function of metallicity
for these tracks. This is overlaid on one of our INT WFC survey fields
(\#76 located  at $\xi  = -1.068^\circ, \eta  = -0.386^\circ$)  from a
central region  of M31 with a  similar colour spread on  the red giant
branch. Evidently there is  good agreement between these theoretical
evolutionary tracks and our M31 data. The INT passbands have been converted to V and I using $I =i'
- 0.101 \times  {\rm (V-I)}$  and ${\rm  V} = V'  + 0.005  \times {\rm
(V-I)}$.  These transformations have been derived by comparison with observations of several Landolt standard fields \footnote{http://www.ast.cam.ac.uk/$\sim$wfcsur/colours.php}.

In  order to  calibrate this  model  to the  data, we  adjust every  I
magnitude for stars in the M31 field using the differential model TRGB
magnitudes with respect to a fiducial colour of ${\rm V-I} = 1.6$.  We
then construct luminosity  functions in the ranges $1.4  < {\rm V-I} <
2.2$ and $2.2 < {\rm V-I} < 2.8$ to represent the metal poor and metal
rich  components respectively.   Using a  data  adaptive least-squares
technique (McConnachie et al.  {\it {in preparation}}), we measure the
relative locations  of the two TRGBs.   The model-corrected metal-rich
TRGB is systematically $\approx$ 0.1 mags brighter than the metal-poor
TRGB and defines an empirical correction to the model-corrected TRGB magnitude.

We now apply this metallicity correction  to field 8 in our CFH data,
as it is both the closest field to M31 and has the highest signal. The
colour magnitude diagram of this field  is shown in the right panel of
Figure \ref{field76}. By applying  the magnitude transformation on the
stars and  by taking into account  the 0.1 mag  difference between the
redder and bluer TRGBs that we would expect were the components all at
one  distance, we conclude  that in  field 8  there is  no significant
difference between  the distance of  the stream component and  that of
the bulk of M31.

\subsection{Relative distance change along the stream}

\begin{figure}
\includegraphics[width=8cm,angle=0]{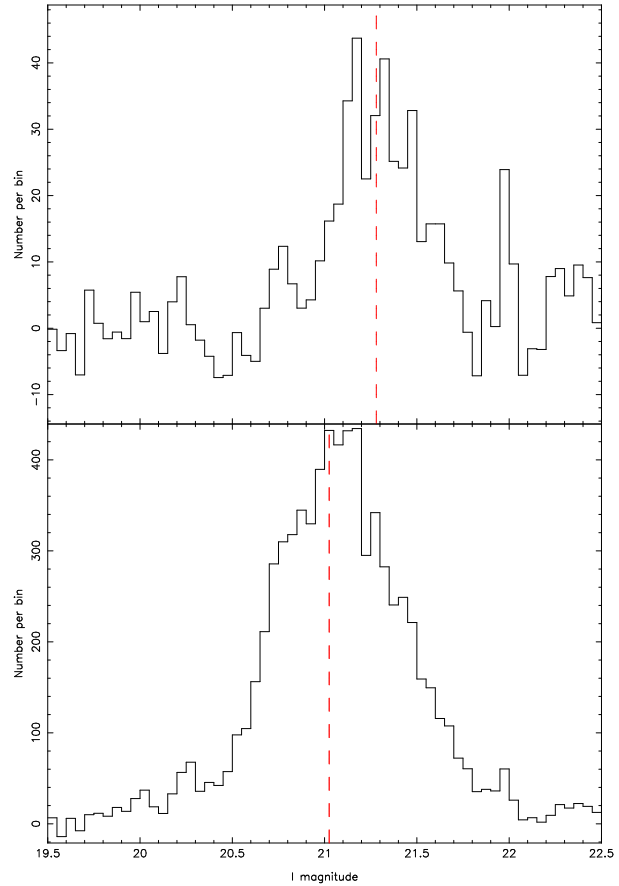}
\caption{The     top      panel     contains     the
foreground-corrected I band luminosity function (using field 14 as the
reference) for field  2, in the range  $2.5 \le V-I \le 3.5$  so as to
isolate  the stream component.  This is  cross correlated  against the
equivalent luminosity  function for field 8 (lower  panel). The dotted
lines are the  locations of the peak in  the luminosity function which
has clearly shifted  to fainter magnitudes for field  2, implying that
the stream here is much further away than in field 8. }
\label{lfshift}
\end{figure}

With a  link between the  stream distance and  the distance of  M31 in
field 8, it is now possible  to examine how this distance changes with
respect  to M31  as a  function  of stream  position.  To  do this  we
cross-correlate  the stream-dominated  I-band luminosity  function ($2.5  < {\rm
V-I} < 3.5$) in each  field with the equivalent luminosity function in
field  8.  Any  shift measured  is a  direct indicator  of  the stream
distance. Examples are shown  in Figure  \ref{lfshift}.   In all
cases field  14 is used as  a reference field and  a scaled (smoothed)
version  of its  luminosity  function is  subtracted  from each  field
analysed  to account  for varying  foreground  Galactic contamination.
The luminosity functions are  constucted using only objects classified
as stellar  sources in  both the  V and I  filters.  Although  for the
redder objects significant incompleteness due to the V-band sets in at
I $\approx$ 22,  this is sufficiently beyond the peak  of the red star
luminosity  function to  not affect  the cross-correlation,  even with
slight  variations  in  field-to-field  depths.  Further  tests  using
I-band only  stellar detections satisfying  ${\rm V-I} >  2.5$ support
this conclusion.

\begin{figure*}
\includegraphics[width=8cm,angle=270]{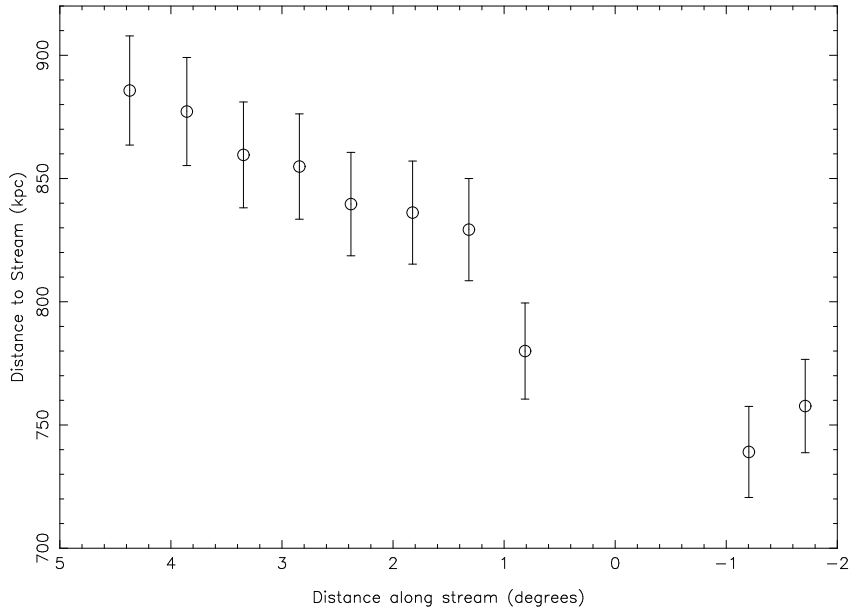}
\caption{The distance  variation along the tidal stream, measured  by cross correlating the  I-band luminosity function
of  field 8  with the  other  fields. The  x-axis is  the distance  in
degrees of the centre of each  field from the intersection of the line
with the semi-major axis of M31. }
\label{distancevar}
\end{figure*}

The   results   of  the   cross-correlation   are   shown  in   Figure
\ref{distancevar} and are tabulated  in Table 1. The main contributors
to the errors in the TRGB distance determinations are: the $rms$ errors
in the  least-squares fit and cross-correlations  which average about
$\pm$0.03  magnitudes;  the photometric  calibration  error of  around
$\pm$0.02 magnitudes; and model-dependent systematics from the various fits of order $\simeq \pm 0.03$. This leads to a total error of around $\pm 20$ kpc in the distances. 

The projection on the sky  and the distance variation along the stream
point to an almost linear  structure in the south-east region, aligned
at  roughly 60  degrees to  the  line of  sight and  extending out  to
distances  well beyond  100 kpc  from the  centre of  M31.   No stream
component is detected in field 14, although there is a clear signal in
the inner  half of  field 13.  The  stream must therefore  be wrapping
around the centre of M31 toward the northeast.  The distance variation
between the two  ends of our survey is  $\approx 140$kpc, implying the
true length of the surveyed region of the stream is $> 160$ kpc.

\begin{table*}
 \begin{minipage}{115mm}
\begin{tabular}{ccccccc}
{}  J2000 & RA & Dec & {$\xi$\footnote{Standard coordinates measured in degrees from the centre of M31.}} &{$\eta ^a$} &{Distance Along Stream\footnote{Distance between the centre of the field and the point of intersection of the resulting line with the semi-major axis of M31.} ($^{\circ}$)} & {Distance to Stream\footnote{Measured relative to M31, assumed to be at a distance of 780kpc.} (kpc)}\\
\hline\\
Field 1 & 0 52 50.36 & 37 17 01.6 & 2.015 & -3.965  & 4.37  & 886 $\pm$ 20 \\
Field 2 & 0 51 32.82 & 37 43 40.9 & 1.745 & -3.525  & 3.86  & 877 $\pm$ 20 \\
Field 3 & 0 50 16.28 & 38 10 12.8 & 1.483 & -3.087  & 3.35  & 860 $\pm$ 20 \\
Field 4 & 0 49 00.39 & 38 36 27.3 & 1.226 & -2.653  & 2.84  & 855 $\pm$ 20 \\
Field 5 & 0 47 43.24 & 38 59 59.5 & 0.969 & -2.264  & 2.38  & 840 $\pm$ 20 \\
Field 6 & 0 46 27.22 & 39 29 51.3 & 0.717 & -1.768  & 1.82  & 836 $\pm$ 20 \\
Field 7 & 0 45 10.35 & 39 56 27.1 & 0.467 & -1.327  & 1.32  & 829 $\pm$ 20 \\
Field 8  & 0 43 53.23 & 40 22 58.1 & 0.219 & -0.886  & 0.81  & 780 $\pm$ 20 \\
Field 12 & 0 38 47.83 & 42 09 18.6 & -0.731 &  0.891  & -1.20 & 739 $\pm$ 20 \\
Field 13 & 0 37 30.49 & 42 36 10.6& -0.963 &  1.342  & -1.71 & 758 $\pm$ 20\\
Field 14 & 0 36 14.95 & 43 02 15.5& -1.186 & 1.781  & --- & --- \\
\hline\\
\end{tabular}
\caption{The Depth Change Along the Stream.}
\end{minipage}
\end{table*}

\section{Conclusions}

The tidal stream  discovered by our survey of M31  (Ibata et al. 2001)
extends linearly  over at  least $6^{\circ}$ of  the sky. In  fact, we
still have not  surveyed out far enough to  find the far south-eastern
end. By analysing the metal  rich red giant branch luminosity function
in  the  I  band and  comparing  them  to  VandenBerg et  al's  (2000)
evolutionary tracks, we  are able to measure a  radial distance change
of order 140  kpc between the two ends of our  survey, with the stream
extending  from approximately 100  kpc behind  to 40  kpc in  front of
M31. We also see evidence that the stream then proceeds to wrap itself
around M31. This  information confirms that the Andromeda  stream is a
gigantic structure,  angled at approximately $60^{\circ}$  to our line
of sight.   Additionally, by analysing the stellar  profile across the
stream we find  that it remains a coherent structure,  at least on the
north-eastern edge. Its stellar distribution also appears to be fairly
constant  with  increasing   galactocentric  distance.  However,  more
observations are  required in  order to determine  the true  extent of
this vast object.

\section * {Acknowledgements}

This  work is  based  on data  collected  at the  Canada-France-Hawaii
Telescope supported  by INSU at Mauna  Kea, Hawaii. The research of AMNF has been supported by a Marie Curie Fellowship of the European Community under contract number HPMF-CT-2002-01758. We would like to thank  Professor VandenBerg  for  supplying us  with his  evolutionary tracks.


\begin{thebibliography} {}

\bibitem[]{} Bellazzini, M., Ferraro, F.R., Ibata, R., 2003, AJ, 125, 188

\bibitem[]{} Carretta, E., Gratton, R.G., Sneden, C., 2000, A\&A, 356,238

\bibitem[]{} Da Costa, G.S., Armandroff, T.E., 1990, AJ, 100, 162

\bibitem[]{} Feltzing, S., Bensby, T., Lundstrom, I., 2003, A\&A, 397L, 1

\bibitem[]{} Ferguson, A. M. N., Irwin, M., Ibata, R., Lewis, G.,Tanvir, N., 2002, AJ, 124, 1452

\bibitem[]{} Ibata, R., Gilmore, G., Irwin, M., 1994, Nature, 370, 194

\bibitem[]{} Ibata, R., Irwin, M., Lewis, G., Ferguson, A. M. N., Tanvir, N., 2001b, Nature, 412, 49 

\bibitem{} Ibata, R., Irwin, M., Lewis, G., Stolte, A., 2001a, ApJ, 551, 294

\bibitem[]{} Irwin, M., Lewis, J., 2001, New AR, 45, 105

\bibitem[]{} Johnston, K.V., Hernquist, L., Bolte, M., 1996, ApJ, 465, 278

\bibitem[]{} Quinn, P.J., Hernquist, L., Fullager, D.P., 1993, ApJ, 403, 74

\bibitem[]{} Sagar, R., Subramaniam, A., Richtler, T., Grebel, E.K., 1999, A\&AS, 135, 391

\bibitem[]{} Salaris, M., Cassisi, S., 1997, MNRAS, 289, 406

\bibitem[]{} Schlegel, D.J., Finkbeiner, D.P., Davis, M., 1998, ApJ, 500, 525 

\bibitem[]{} Schwarzkopf, U., Dettmar, R.-J., 2000, A\&A, 361, 451 

\bibitem[]{} VandenBerg, D.A., Swenson, F.J., Rodgers, F.J., Iglesias, C.A., Alexander, D.R., 2000, ApJ, 532, 430

\bibitem[]{} Wheeler,J.C., Sneden, C., Truran, J.W., 1989, ARA\&A, 27, 279 

\bibitem[]{} White, S.D.M., Rees, M.J., 1978, MNRAS, 183, 341

\end{thebibliography}
\end{document}